\documentclass[letterpaper,twocolumn,english,prl,showpacs]{revtex4-1}
\usepackage{graphicx}
\usepackage{bm}
\usepackage[linkcolor=blue]{hyperref}
\usepackage{amsmath,amssymb}
\usepackage{subfigure}
\usepackage{lipsum}
\usepackage{color}
\usepackage{hyperref}
\usepackage{footnote}

\begin{document}

\title{Monolithic Semiconductor Chips as a Source for Broadband Wavelength-Multiplexed Polarization Entangled Photons}
\author{Dongpeng Kang}
\email{dongpeng.kang@mail.utoronto.ca}
\affiliation{The Edward S. Rogers Department of Electrical and Computer Engineering, Centre for Quantum Information and Quantum Control,
University of Toronto, 10 King\textquoteright{}s College Road, Toronto,
Ontario M5S 3G4, Canada.}

\author{Ankita Anirban}
\affiliation{The Edward S. Rogers Department of Electrical and Computer Engineering, Centre for Quantum Information and Quantum Control,
University of Toronto, 10 King\textquoteright{}s College Road, Toronto,
Ontario M5S 3G4, Canada.}

\author{Amr S. Helmy}
\affiliation{The Edward S. Rogers Department of Electrical and Computer Engineering, Centre for Quantum Information and Quantum Control,
University of Toronto, 10 King\textquoteright{}s College Road, Toronto,
Ontario M5S 3G4, Canada.}

\date{\today }

\begin{abstract}
Generating entangled photons from a monolithic chip is a major milestone towards real-life applications of optical quantum information processing including quantum key distribution and quantum computing.  
Ultrabroadband entangled photons are of particular interest to various applications such as quantum metrology and multi-party entanglement distribution. In this work, we demonstrate the direct generation of broadband wavelength multiplexed polarization entangled photons from a semiconductor chip for the first time. Without the use of any off-chip compensation, interferometry, entangled photons with a signal-idler separation as large as 95 nm in the telecom band were observed. The highest concurrence of 0.98$\pm$0.01 achieved in this work is also the highest, to the best of our knowledge, comparing to all previously demonstrated semiconductor waveguide sources. 
This work paves the way for fully integrated, ultrabroadband sources of polarization entangled photons.
\end{abstract}

\maketitle

\par Entangled photons play pivotal roles in quantum information processing (QIP). They are essential sources for optical quantum computing (QC) \cite{Ladd_Nature_2010}, photon-based quantum metrology (QM) \cite{Giovannetti_NP_2011} and quantum key distribution (QKD) \cite{Gisin_RMP_2002}. Each application calls for entangled photon pairs with specific attributes. Interest in broadband entangled photon pairs has increased in recent years. This has been motivated by various applications in quantum optical technologies such as quantum optical coherence tomography (QOCT) \cite{Nasr_PRL_2003} and quantum optical frequency comb sources (QOFC) \cite{Chen_PRL_2014,Roslund_NP_2014}. In QOCT, the use of ultrabroadband photon pairs with extremely short correlation times can enhance the accuracy of protocols for quantum positioning and timing beyond the classical limits \cite{Nasr_PRL_2003}. In the regime of continuous variable (CV) entanglement, QOFC, generated by placing a broadband spontaneous parametric down-conversion (SPDC) crystals in an optical parametric oscillator (OPO) cavity, is a crucial resource for multi-mode entanglement and cluster states \cite{Chen_PRL_2014}. Spectroscopy modalities that are enabled by entangled photon sources also rely on the availability of such photons within a broad range of energies to match transitions of various material systems \cite{Mukamel_NatComm_2013}.

\par One emerging application of entangled photons is in the multi-party entanglement distribution networks \cite{Chapuran_NJP_2009}, which can empower multi-user QKD systems and quantum communication networks. In this case, broadband entangled photon pairs are required by the service provider who, by virtue of the relevant multi-user QKD protocols is required to generate multiple pairs of entangled photons at virtually the same time and subsequently distribute paired photons to multiple users at different locations. To realize such entanglement distribution networks, one practical methodology involves utilizing existing fiber communication links and dense wavelength-division-multiplexing (DWDM) systems to re-route entangled photon pairs that possess slightly different wavelengths to users at different locations. As such, the service provider for these systems requires the availability of a vast array of entangled photon sources at different wavelengths. Given the form-factor of existing single channel entangled photon sources, the scalablity of these systems can be greatly enhanced if a single source can generate all the required entangled photon pairs \cite{Lim_OFT_2010}. There are no existing approaches or technology, which are able to address this unmet demand. 

\par The generation of wavelength-multiplexed entangled photons requires a combination of techniques for producing both broad bandwidth and polarization entanglement. Typical photon pairs generated via type-II SPDC processes have a narrow bandwidth of only a few nanometers \cite{Kaiser_NJP_2012,Herrmann_OE_2013}, due to the material birefringence, and therefore are not suitable for this application because of the insufficient use of the fiber communication bandwidth. Broadband photon pairs generated via type-I or type-0 processes can be used to generate wavelength-multiplexed entangled photons, however, only with complex interferometric setups due to the lack of orthogonal polarizations from a single process \cite{Lim_OE_2008,Herbauts_OE_2013,Arahira_OE_2013}. For practical applications, compact, robust sources of entangled photons capable of integrating with other components in the system should be developed. Recent developments in integrated quantum photonics have achieved compact generation of entangled photons in monolithic waveguide devices using SPDC \cite{Orieux_PRL_2013,Horn_SR_2013,Kang_PRA_2015} or spontaneous four-wave mixing (SFWM) \cite{Matsuda_SR_2012,Olislager_OL_2013,Lv_OL_2013}. However, the demonstration of broadband wavelength-multiplexed entangled photons based on a single waveguide chip has not been reported. In addition, these existing semiconductor sources all suffer from low degree of entanglement.

\par In this work, we demonstrate a single-chip broadband wavelength multiplexed polarization entangled photon source for the first time. The source is a modal phase matched Bragg reflection waveguide (BRW) \cite{Horn_SR_2013}, consisting of multiple layers of AlGaAs, in which the pump light is guided by Bragg reflections from two periodic reflectors while the down-converted photons are guided by conventional total internal reflections. Because AlGaAs is non-birefringent, cross-polarized photons generated from the type-II process have a much broader spectrum than those generated from birefringent crystals, allowing for wavelength multiplexing using standard DWDM. In addition, due to the lack of birefringence, the down-converted photons propagate at almost the same group velocities, which enables the generation of wavelength multiplexed polarization entangled photons directly from the chip without any off-chip compensation or interferometry. The degree of entanglement is also the highest to date among all semiconductor waveguide sources, to the best of our knowledge. 
Based on III-V semiconductors, this source allows monolithic integration of its own pump laser, which can lead to electrically pumped, room-temperature operating sources of broadband wavelength multiplexed polarization entangled photons.

\section{Results}
\subsection{Theory: Broadband entanglement in Bragg reflection waveguides}
\par BRWs have been extensively studied in the past few year for classical wavelength conversions \cite{Abolghasem_JSTQE_2012,Abolghasem_OL_2014} and entangled photon pair generations \cite{Horn_SR_2013,Kang_OL_2012,Kang_PRA_2015}. The two distinct guiding mechanisms involved provide a great design flexibility in achieving phase matching near the bandgap of semiconductors where formidable material dispersion exists, as well as in generating photon pairs with particulate quantum properties, such as entanglement \cite{Kang_OL_2012,Zhukovsky_PRA_2012,Kang_PRA_2014} and spectral separability \cite{Svozilik_OE_2011}. 

\par Of particular interest to this work, it was proposed that dispersion engineered BRWs with zero group velocity mismatch (GVM) between photons in a pair can generate ultra-broad band wavelength multiplexed polarization entangled photons \cite{Svozilik_OE_2012}. However, even without any waveguide dispersion engineering which sacrifices the generation efficiency, cross-polarized photon pairs generated via a type-II SPDC process in a BRW showed inherent polarization entanglement without any off-chip compensation, usually required to remove the walk-off between photons in a pair, due to the lack of material birefringence \cite{Horn_SR_2013}. As such, it can be expected that the photon pairs have a much broader bandwidth than a typical type-II process in birefringent materials because the bandwidth is largely determined by the GVM between photons in a pair.
\par
\begin{figure}[t]
\centering
\includegraphics[width=0.95\columnwidth]{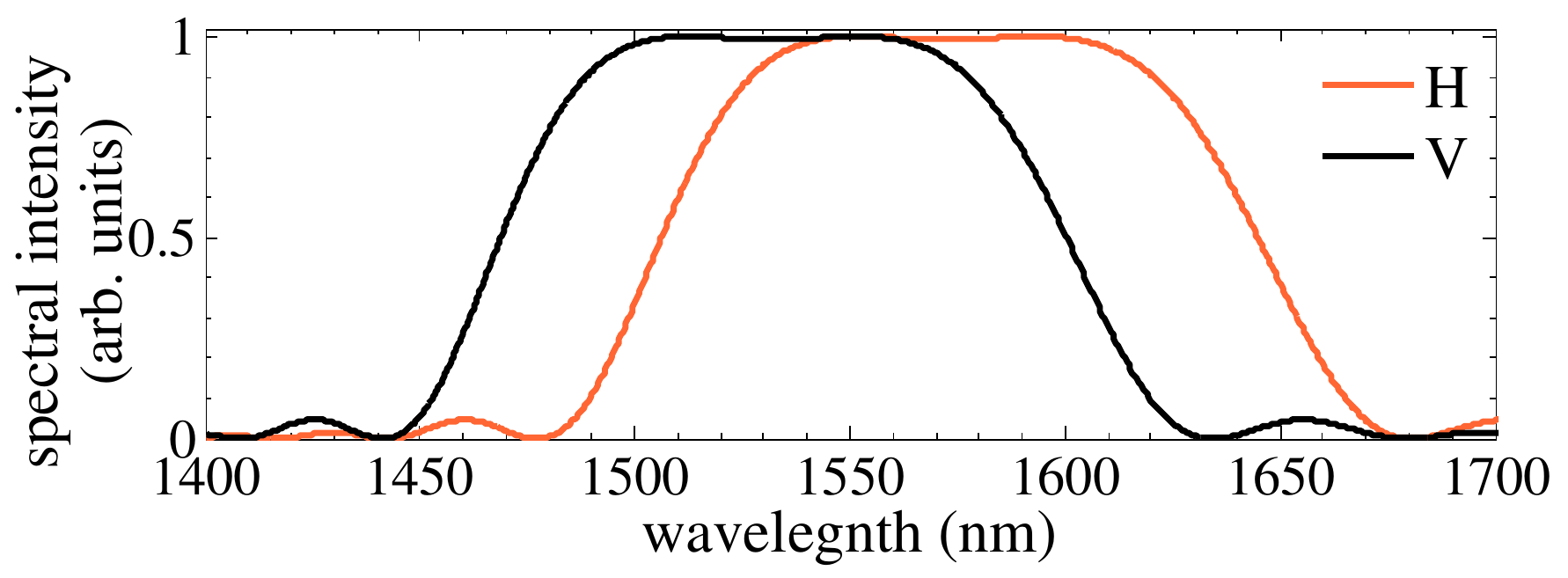}
\caption{The simulated spectra for the H and V polarized SPDC photons from a 1.09 mm long BRW using a CW pump.}
\label{Fig:spectra}
\end{figure}

 We consider the same BRW demonstrated for polarization entanglement generation in \cite{Horn_SR_2013}, with the detailed structure described in Methods. For a 1.09 mm long waveguide in this work, the simulated spectra for the horizontal (H) and vertical (V) polarized down-converted photons using a CW pump are shown in Fig. \ref{Fig:spectra}. The spectrally overlapped region of the two polarizations has a FWHM of $\sim$95 nm, which is considerably wider than that of a typical type-II process in a birefringence material.
\begin{figure}[t]
\centering
\includegraphics[width=0.99\columnwidth]{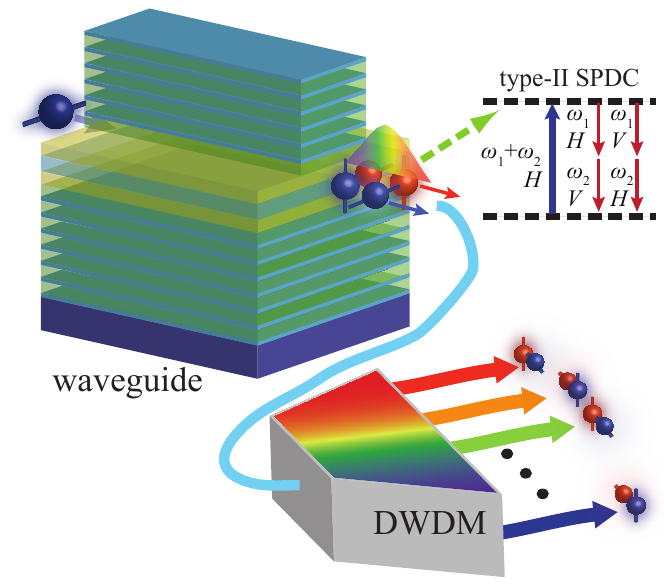}
\caption{The schematic setup of the thought experiment. Broadband polarization entangled photons are generated via a type-II SPDC process in the waveguide. They are subsequently coupled into an optic fiber and sent to a DWDM to be deterministically separated into many wavelength channel pairs.}
\label{Fig:schematic_setup}
\end{figure}
\par Now consider the experiment that a DWDM filter is used after the waveguide to deterministically separate the down-converted photons into many wavelength channel pairs, as schematically shown in Fig. \ref{Fig:schematic_setup}. The output two-photon state of each channel pair is given by
\begin{align}
\left\vert \psi_n\right\rangle =&\frac{1}{\sqrt{2}}\iint_{B_n}{\mathrm{d}\omega _{s}\mathrm{d}\omega _{i}}[\Phi_{HV}(\omega_s,\omega_i)a_{H}^{\dag }\left( \omega _{s}\right)a_{V}^{\dag }\left(\omega _{i}\right) \nonumber \\ 
&+\Phi_{VH}(\omega_s,\omega_i)a_{V}^{\dag }\left( \omega _{s}\right)a_{H}^{\dag }\left( \omega _{i}\right)]\left\vert vac\right\rangle,
\label{Eq:state}
\end{align}
where $\omega_s$, $\omega_i$ represent the signal, idler channel frequencies; $\Phi_{VH}(\omega_s,\omega_i)$, $\Phi_{HV}(\omega_s,\omega_i)$ are the associated join spectral amplitudes that the signal photon is vertically (V) polarized and idler photon is horizontally (H) polarized, and vice versa. The integration is performed over the channel pair transmission bands $B_n$ of the DWDM. The state given by Eq. (\ref{Eq:state}) is maximally entangled in polarization, as long as the spectral-temporal indistinguishability is satisfied in the two term on the RHS of Eq. (\ref{Eq:state}). According to Fig. \ref{Fig:spectra}, the spectral intensities of H and V photons are almost identical near degeneracy. In addition, the temporal walk-off between the two polarizations due to modal birefringence is negligible comparing to the photon coherence time. As a result, high degree of entanglement can be achieved for channel pairs near the degenerate wavelength.
\begin{figure}[t]
\centering
\includegraphics[width=0.99\columnwidth]{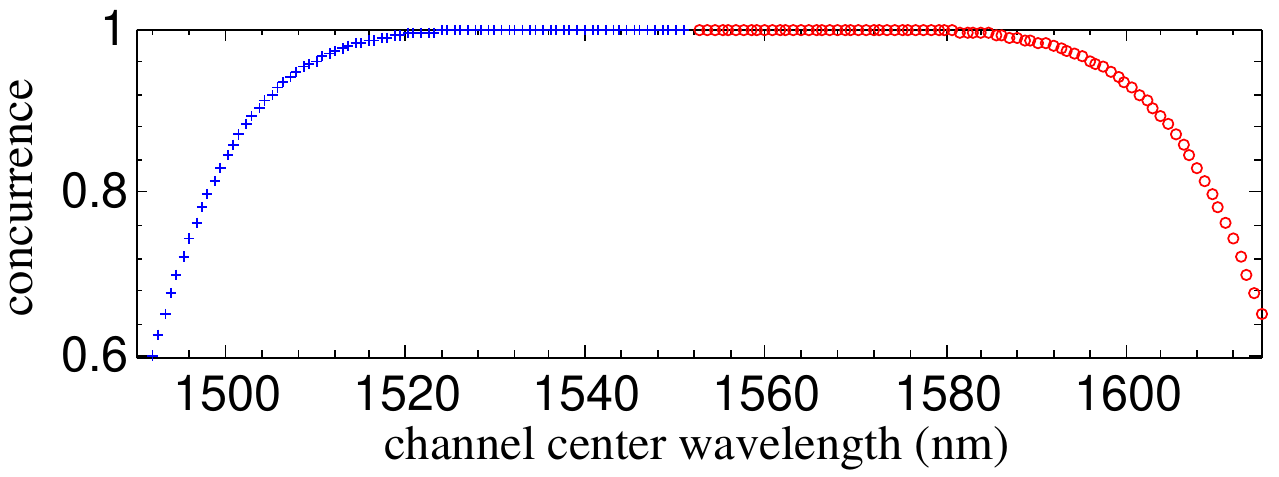}
\caption{The calculated concurrence as a function of signal and idler wavelengths, shown in different colours, for polarization entangled photons generated from the BRW in consideration.}
\label{Fig:concurrence_vs_wavelength}
\end{figure}
\par Assuming a typical 55 GHz (0.44 nm) channel bandwidth and 100 GHz (0.8 nm) channel separation for the DWDM, the calculated concurrence \cite{Wootters_PRL_1998}, an entanglement monotone, as a function of signal and idler wavelengths, is shown in Fig. \ref{Fig:concurrence_vs_wavelength}. According to Fig. \ref{Fig:concurrence_vs_wavelength}, the concurrence is almost 1 near the degenerate wavelength, indicating a maximal degree of entanglement.  A concurrence of at least $1/\sqrt{2}\sim 0.71$, the minimal value that guarantees Clauser-Horne-Shimony-Holt (CHSH) inequality violation \cite{Verstraete_PRL_2002}, could be generated within a range of $\sim$118 nm of wavelength. This corresponds to over 70 pairs of frequency channels. As a result, this BRW chip can generate over 70 channel pairs of polarization entangled photons directly, with a concurrence of at least 0.71 in the ideal case.
\par The bandwidth is limited by the residual form birefringence of the waveguide. To further increase the bandwidth, one strategy is to cancel the GVM between the cross-polarized down-converted photons via dispersion engineering, as suggested in \cite{Zhukovsky_PRA_2012,Svozilik_OE_2012}. Alternatively, the bandwidth can be increased by tapering the waveguide, which creates a gradient of phase mismatch along the direction of propagation, in a close analogy to chirping in quasi-phase matching (QPM) \cite{Nasr_PRL_2008}. In addition, ultrabroadband polarization entanglement can be achieved in BRWs via concurrent type-0 and type-I SPDC processes \cite{Kang_OL_2012,Kang_PRA_2015}. In such a case, co-polarized photons in a pair naturally exhibit broader spectral bandwidths than in a type-II process. 
\subsection{Experimental Demonstration}
\par
\begin{figure}[t]
\includegraphics[width=0.99\columnwidth]{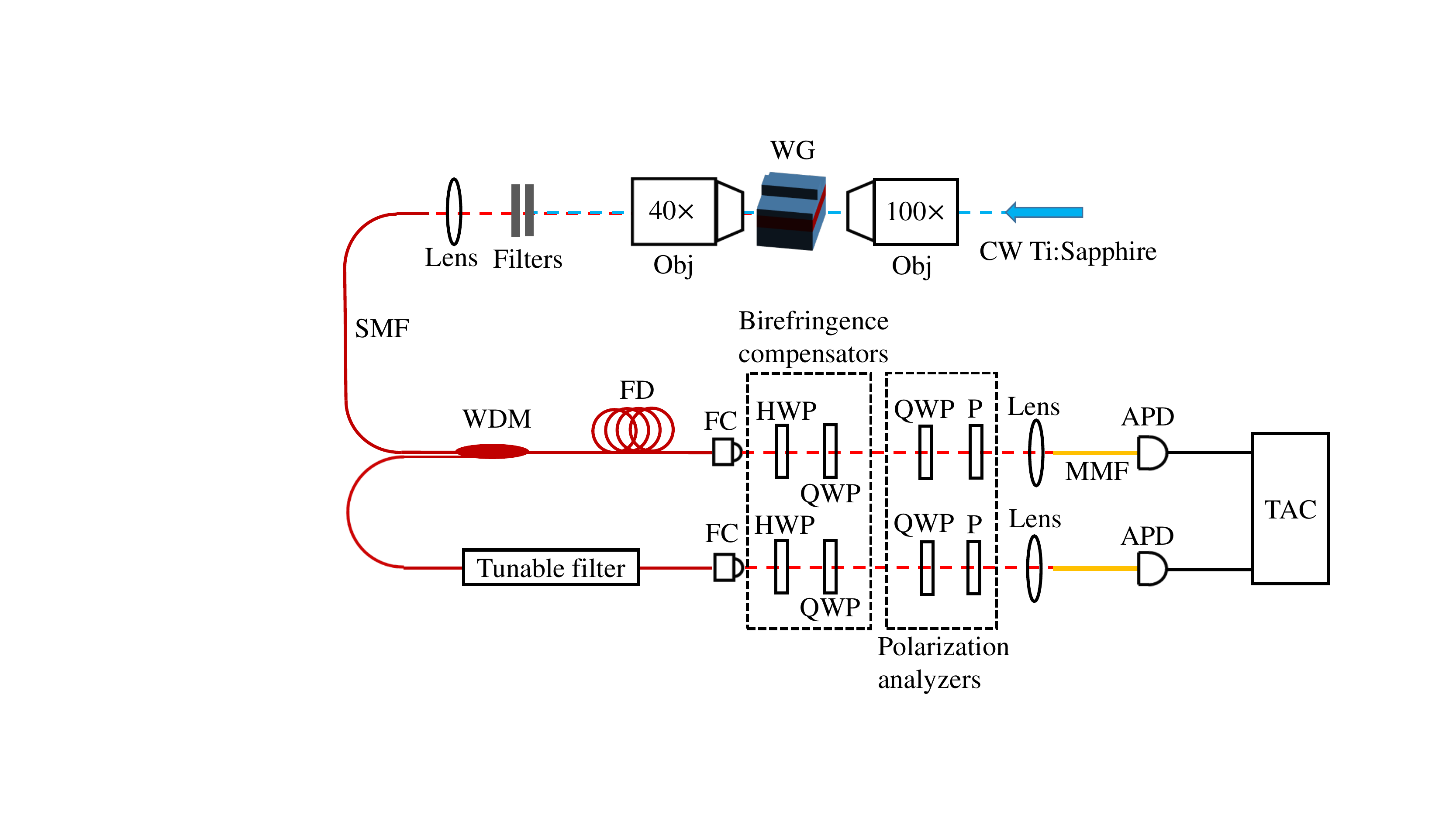}
\caption{The schematic of the experimental setup. Obj: objective; WG: waveguide; SMF: single-mode fiber; WDM: wavelength division multiplexer filter; FD: fiber delay; FC: fiber collimator; HWP: half-wave plate; QWP: quarter-wave plate; P: polarizer; MMF: multimode fiber; APD: InGaAs avalanche photo-diode; TAC: time-to-amplitude converter.}
\label{Fig:setup}
\end{figure}

\begin{figure}[t]
\includegraphics[width=0.99\columnwidth]{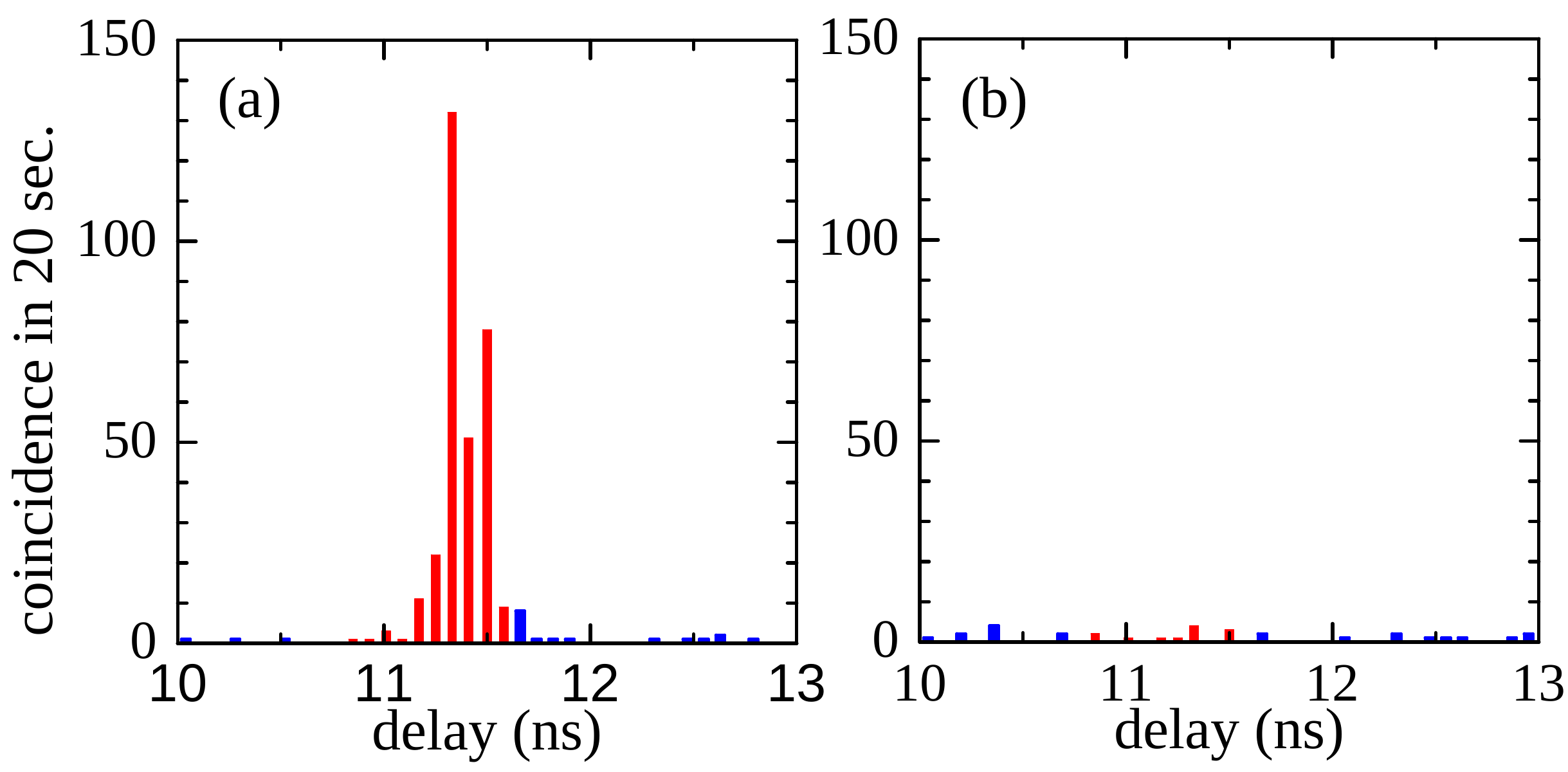}
\caption{Coincidence histograms for a channel pair near the degenerate wavelength in the (a) $D-D$, and (b) $D-A$ basis, respectively. The red color represents the coincidence counts in a coincidence window of $\sim$0.8 nm.}
\label{Fig: histogram}
\end{figure}

 To demonstrate the theoretical predictions presented above, we tested a 1.09 mm long waveguide sample for the experiment proposed in Fig. \ref{Fig:schematic_setup}. The experimental setup is shown in Fig. \ref{Fig:setup} and described in Method. Examples of coincidence histograms in the diagonal basis are given in Fig. \ref{Fig: histogram}. Near the degenerate wavelength, the generation rate is estimated to be $1\times 10^4$ pairs/s/mW/GHz in terms output photon pairs after the waveguide and internal pump power.

\par
\begin{figure}[t]
\centering
\includegraphics[width=0.99\columnwidth]{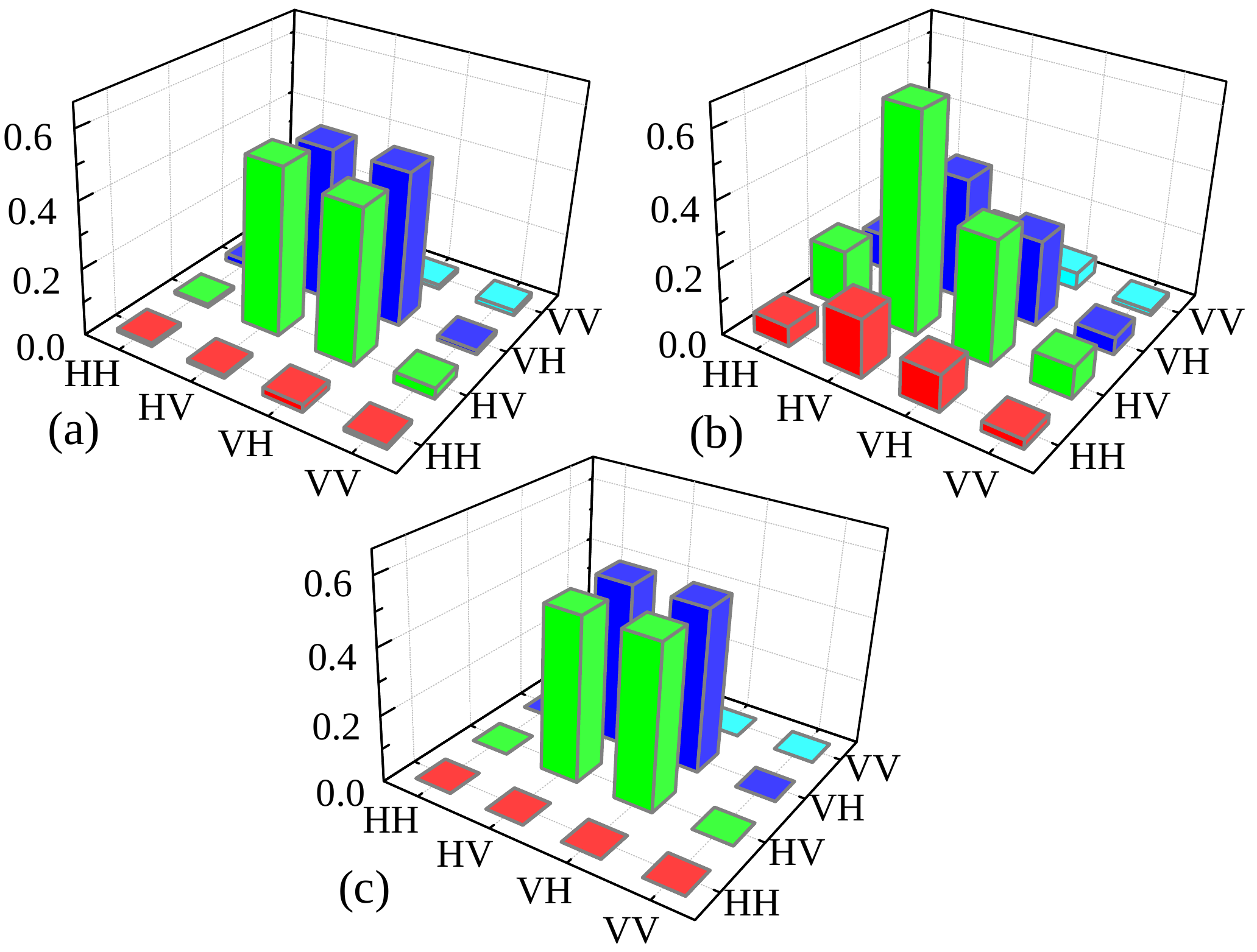}
\caption{The amplitudes of the reconstructed net density matrices for channel pairs of (a) 1554.0-1557.4 nm and (b) 1509.7-1604.6 nm, as well as (c) that of the maximally entangled state.}
\label{Fig: densitymatrix_WDM}
\end{figure}

 Quantum state tomography for each channel pair was performed by measuring the coincidence counts in 16 polarization settings (H, V, D, R in each channel), and density matrices were reconstructed using the maximum-likelihood method \cite{James_PRA_2001}. As examples, the amplitudes of the density matrices, after subtracting the accidentals, are shown in Fig. \ref{Fig: densitymatrix_WDM}(a) and (b) for 1554.0-1557.4 nm channel pair close to degeneracy, and 1509.7-1604.6 nm channel pair away from degeneracy, respectively. For a direct comparison, that of the maximally entangled state $(\left|HV\right\rangle+\exp{(i\phi)}\left|VH\right\rangle)/\sqrt{2}$ is shown in Fig. \ref{Fig: densitymatrix_WDM}(c). 
\par The density matrices near the degenerate wavelength, such as the one in Fig. \ref{Fig: densitymatrix_WDM}(a), show high resemblances with that of a maximally entangled state in Fig. \ref{Fig: densitymatrix_WDM}(c), indicating a high degree of entanglement. When the wavelengths are significantly far from degeneracy, such as in \ref{Fig: densitymatrix_WDM}(b), the available photon pairs decrease, along with the decrease of entanglement.

\par
\begin{figure}[t]
\centering
\includegraphics[width=0.99\columnwidth]{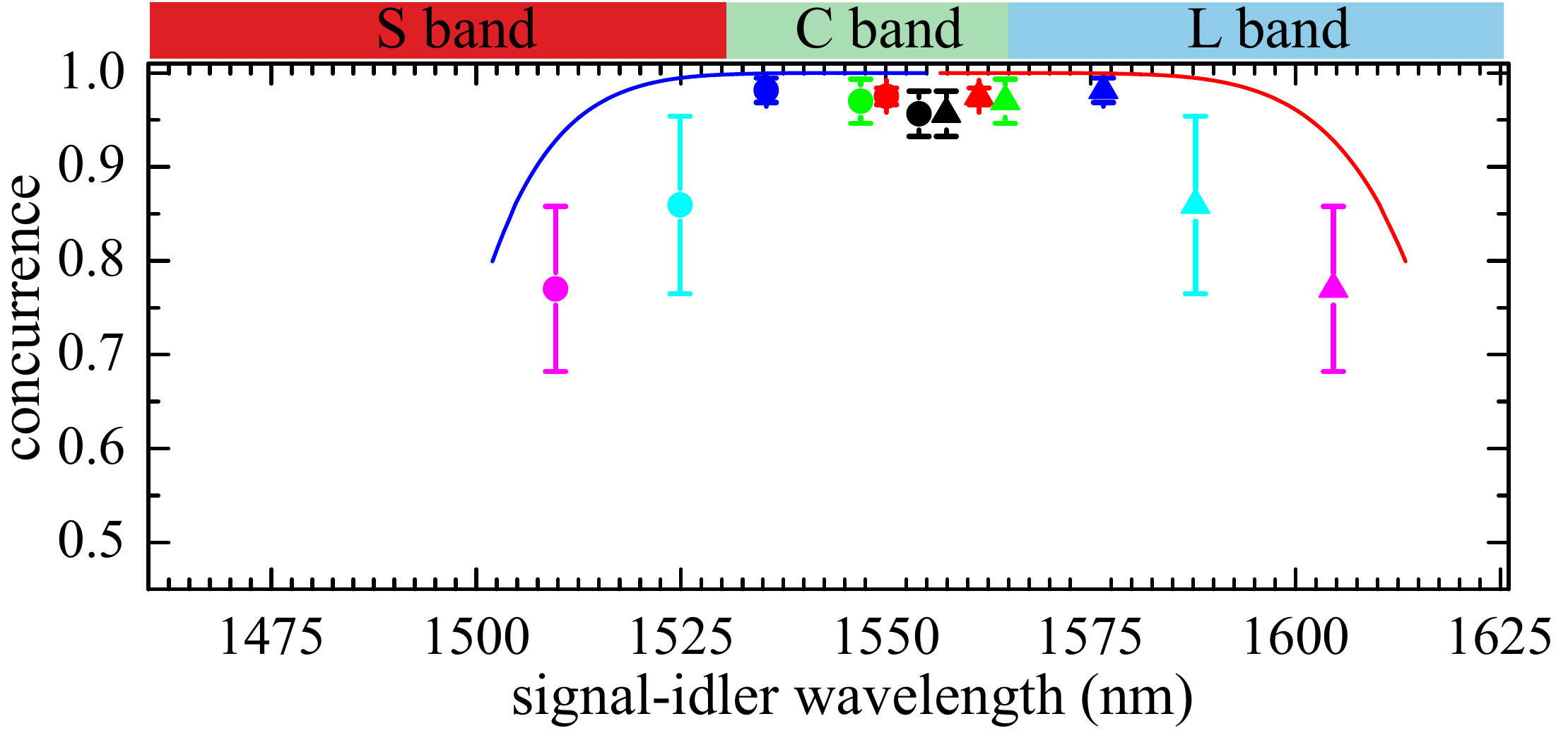}
\caption{The dependence of concurrence on the signal and idler wavelengths. Wavelength channels in the same pair are represented by the same color. The solid curves are the theoretical predictions from Fig. \ref{Fig:concurrence_vs_wavelength}, with the degenerate wavelength shifted 6 nm to match that in the experiment.} 
\label{Fig:con_vs_wavelength_with_theory}
\end{figure}

 The results for all measured channel pairs are summarized in Fig. \ref{Fig:con_vs_wavelength_with_theory}, which shows the concurrence calculated from the measured density matrix as a function of the signal and idler wavelengths. A high value of concurrence of at least $0.96\pm0.02$ and as high as $0.98\pm0.01$, as well as a highest fidelity of 0.97 to a maximally entangled state $(\left|HV\right\rangle+\exp{(i\phi)}\left|VH\right\rangle)/\sqrt{2}$ could be observed for the signal-idler pairs from 1535.4 nm to 1575.6 nm, which approximately covers the whole C-band (1530 nm-1565 nm). This corresponds to a total bandwidth of over 40 nm for the down-converted photons, and equivalently 26 pairs of wavelength channels if the channel spacing is 100 GHz. For signal and idler wavelengths further apart, the concurrence starts to decrease as well as the coincidence counts. Nevertheless, a concurrence of $0.77\pm0.09$ could still be measured for the channel pair of 1509.7-1604.6 nm, with a wavelength separation of 95 nm. Note in S- and L-bands, the increased error bars could be due to the decreased coincidence counts. The concurrences being lower than expected could also be because of the errors in estimating the waveguide dispersion properties, which determine the bandwidth of photon pairs.

\par The measured entanglement bandwidth extends from S- to L-bands, which can enable a high utilization of existing fiber optic communication infrastructures from a single source. 
We emphasize that polarization entanglement is generated directly from the monolithic chip, without any off-chip compensation or interferometer, comparing with existing waveguide sources. The degree of entanglement is also the highest reported in all semiconductor waveguide sources of polarization entangled photons, as summarized in Table \ref{Table:comparison}. 
%
\begin{table}[tb]
\caption{Comparison of degree of entanglement generated from all demonstrated monolithic waveguide sources of polarization entangled photons.}
\begin{tabular}{l c c c}\\
\hline
Reference & Platform & Concurrence & Fidelity \\ \hline
\cite{Orieux_PRL_2013} & AlGaAs & 0.75$\pm$0.05 &  0.87$\pm$0.03\\
\cite{Horn_SR_2013} & AlGaAs &  0.52 & 0.83 \\
\cite{Kang_PRA_2015} & AlGaAs & 0.85$\pm$0.07 & 0.89 \\
\cite{Matsuda_SR_2012} & silicon & 0.88$\pm$0.02 & 0.91$\pm$0.02 \footnote{The concurrence and fidelity in \cite{Matsuda_SR_2012} are raw values without noise subtraction. As a comparison, the highest raw concurrence and fidelity obtained in this work are 0.92$\pm$0.03 and 0.95, respectively. Results from all other references are noise subtracted.}\\
\cite{Olislager_OL_2013} & silicon & NA & 0.87 \\ 
This work & AlGaAs & $0.98\pm0.01$ & 0.97 \\\hline
\end{tabular}
\label{Table:comparison}
\end{table}

\par
\begin{figure}[t]
\centering
\includegraphics[width=0.99\columnwidth]{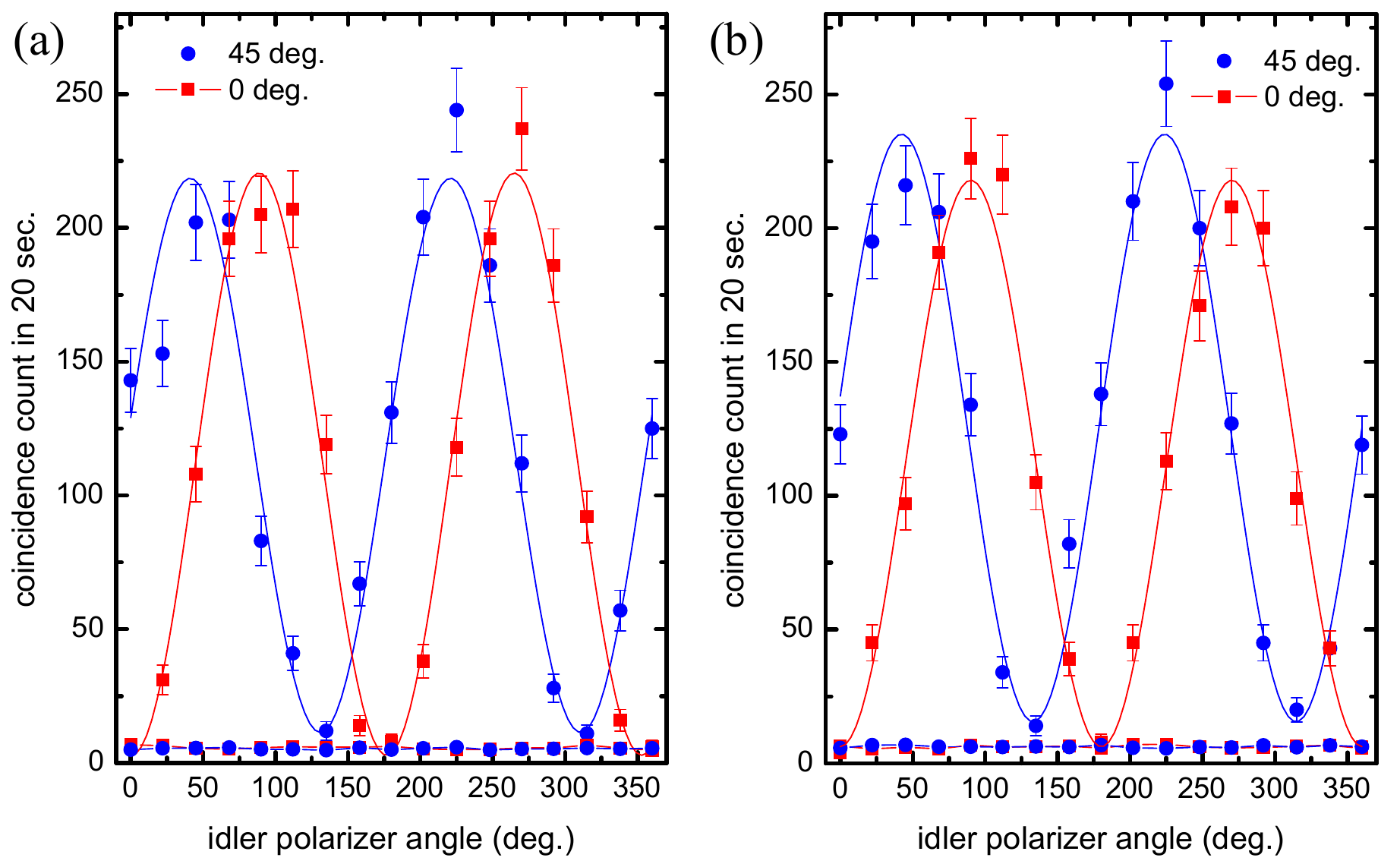}
\caption{Interference measurements for channel pairs (a) 1550.9-1561.8 nm and (b) 1547.0-1565.8 nm in the H and D basis. The solid curves are the sinusoidal fits. Accidental counts in each basis are also shown.}
\label{Fig:SH3_WDM_vis}
\end{figure}

 Lastly, we show the quantum interference in a Bell-type experiment for a few channel pairs near the degenerate wavelength in H and D basis, in which the polarization of signal channel was set to be 0$^\circ$ and 45$^\circ$, respectively, while the polarization of the idler channel was varied. Examples of interference patterns are given in Fig. \ref{Fig:SH3_WDM_vis}(a) for channel pair 1550.9-1561.8 nm and Fig. \ref{Fig:SH3_WDM_vis}(b) for channel pair 1547.0-1565.8 nm. The fitted curves, without accidental subtraction, show visibilities of 97.3\% and 94.5\% in the H basis, and 90.1\% and 87.5\% in the D basis, respectively.

\section{Discussion}

\par In summary, we demonstrated the generation of broadband wavelength multiplexed polarization entangled photons directly from a monolithic waveguide chip for the first time. With a minimal concurrence of 0.77, polarization entanglement was demonstrated in a bandwidth of 95 nm. In particular, for over 40 nm near the degenerate wavelength, high quality polarization entangled photons with a minimal concurrence of 0.96 and as high as 0.98 can be achieved.  In addition, the degree of entanglement demonstrated in this work is the highest reported among all monolithic waveguide sources to the best of our knowledge.

\par The performance of the source can be improved by better device fabrications that decrease waveguide losses. As predicted in \cite{Zhukovsky_JOSAB_2012}, the generation rate can be increased by orders of magnitude in a lossless waveguide. This will allow one to afford using a DWDM with a narrower channel bandwidth and closer channel separation. The degree of entanglement can be improved due to the increased spectral-temporal indistinguishability. On the other hand, the entanglement bandwidth could be increased via dispersion engineering, tapering, or by utilizing concurrent type-0 and type-I processes, as mentioned before.

\par Comparing with existing techniques for generating broadband wavelength multiplexed entangled photons, this method requires no interferometry, and thus greatly reduced the source complexity. Such a chip form-factor is essential for real life applications of entangled photons. In addition, based on a III-V semiconductor platform, we can envisage the integration of its own electrically pumped laser and passive components such as an array waveguide grating (AWG) on the same chip, which paves the way for fully integrated sources in quantum information processing and entanglement networks.

\section{Method}
\subsection{Sample description}

\par The semiconductor wafer was grown on a GaAs [001] substrate using metal–organic chemical vapor deposition (MOCVD). The composite core consists of a 500 nm thick Al$_{0.61}$Ga$_{0.39}$As sandwiched by a pair of 375 nm Al$_{0.20}$Ga$_{0.80}$As. Each of the two Bragg mirrors consists of 6 pairs of 129/461 nm Al$_{0.25}$Ga$_{0.75}$As/Al$_{0.70}$Ga$_{0.30}$As. The waveguide used for polarization entanglement measurements has an etch depth of 3.6 $\mu$m and a ridge width of 4 $\mu$m. The propagation losses for the down-converted photons were measured to be $\sim$4 cm$^{-1}$.

\subsection{Experimental details}

\par The sample was mounted on a standard end-fire setup, with the pump from a cw Ti:sapphire laser set at TE polarization and the degenerate wavelength of 777.86 nm. A 100$\times$ objective lens was used to focus the pump beam into the Bragg mode of the waveguide. After the sample, the output light was collected by a 40$\times$ objective lens and filtered by a few long pass filters to reject the pump, and then coupled into a single mode fiber. A fiber based dichroic splitter was then subsequently used to deterministically separate paired photons into signal and idler paths according to wavelengths. One of the paths then subsequently passed through a tunable bandpass filter with a bandwidth set to be 55 GHz. This can select out paired photons in a given frequency channel pairs. Note that, ideally, an additional bandpass filter should be used in the other path to select the other twin photon in the same pair and suppress the uncorrelated noise photons. This could significantly reduce the accidental count rate while maintaining a high coincidence. Instead, a commercial DWDM could also be used to replace the filter system used here. Nevertheless, our filters emulates a DWDM in terms of the physics and allows us to explore the maximum bandwidth achievable for the device itself. 

\par 
Photons in the two frequency channels were then launched into free-space for polarization manipulation and analyzing before being detected by a pair of InGaAs single photon detectors. The photons from the narrow frequency channel were detected by a free-running detector (id220, ID Quantique, 20\% quantum efficiency), which triggered a gated detector (id210, ID Quantique, 25\% quantum efficiency) used to detect photons from the wide frequency channel. An extra length of fiber was connected before the gated detector in order to compensate for the electronic delay between the two detectors. The detection events from both detectors were then sent to a time-to-amplitude converter (TAC) (id800, ID Quantique) with a timing resolution of $\sim80$ ps to record the coincidence histograms. 

\par Near the degenerate wavelength, the peak net coincidence counts for all channels were approximately 350 counts in 20 seconds under an external pump power of 15 mW before the input objective lens. The pump power and integration time were chosen as a compromise of the coincidence to accidental (CAR) rate and the total measurement time, due to the mechanic instability of the experimental equipment. The peak CAR were $\sim$50 in this case. Taking into account the pump coupling efficiency (7.3\%), overall photon losses due to the output objective lens, fiber coupling, filters, wave-plates and polarizers, as well as detector efficiencies, we estimate a generation rate of $1\times 10^4$ pairs/s/mW/GHz in terms output photon pairs after the waveguide and internal pump power. For two-photon interference measurements shown in Fig. \ref{Fig:SH3_WDM_vis}, the external pump power was set to be 10 mW, when the stability of the setup can be sustained for a sufficiently long duration to carry out all the measurements.

\par To measure the polarization of each photon, a pair of birefringence compensators consisting of a half-wave plate (HWP) and a quarter-wave plate (QWP) was inserted in each path to partially compensate for the polarization rotation in the fiber, such that a given polarization before the detectors corresponds to the same polarization immediately after the waveguide. Projective polarization measurements were then performed using a QWP and a polarizer in each path.

\section*{Funding Information}
Natural Sciences and Engineering Research Council of Canada (NSERC).

\section*{Acknowledgments}

The authors would like to thank Eric Y. Zhu and Feihu Xu for helpful discussions.

\end{document}